\documentclass[9pt,twocolumn,twoside]{pnas-new}
\templatetype{pnasinvited} % Choose template 
\definecolor{CLBlue}{rgb}{0, .25, .8}

\title{How humans learn and represent networks}

\author[a]{Christopher W. Lynn}
\author[a,b,c,d,e,f,1]{Danielle S. Bassett} 

\affil[a]{Department of Physics \& Astronomy, College of Arts \& Sciences, University of Pennsylvania, Philadelphia, PA 19104, USA}
\affil[b]{Department of Bioengineering, School of Engineering \& Applied Science, University of Pennsylvania, Philadelphia, PA 19104, USA}
\affil[c]{Department of Electrical \& Systems Engineering, School of Engineering \& Applied Science, University of Pennsylvania, Philadelphia, PA 19104, USA}
\affil[d]{Department of Neurology, Perelman School of Medicine, University of Pennsylvania, Philadelphia, PA 19104, USA}
\affil[e]{Department of Psychiatry, Perelman School of Medicine, University of Pennsylvania, Philadelphia, PA 19104, USA}
\affil[f]{Santa Fe Institute, Santa Fe, NM 87501, USA}

\leadauthor{Lynn} 

\authorcontributions{C.W.L. and D.S.B. conceived the project, C.W.L. wrote the paper, and D.S.B. edited the paper.}
\authordeclaration{The authors declare no competing financial interests.}
\correspondingauthor{\textsuperscript{1}To whom correspondence should be addressed. E-mail: dsb@seas.upenn.edu.}


\begin{abstract}
Humans receive information from the world around them in sequences of discrete items -- from words in language or notes in music to abstract concepts in books and websites on the Internet. In order to model their environment, from a young age people are tasked with learning the network structures formed by these items (nodes) and the connections between them (edges). But how do humans uncover the large-scale structures of networks when they only experience sequences of individual items? Moreover, what do people's internal maps and models of these networks look like? Here, we introduce \textit{graph learning}, a growing and interdisciplinary field studying how humans learn and represent networks in the world around them. Specifically, we review progress toward understanding how people uncover the complex webs of relationships underlying sequences of items. We begin by describing established results showing that humans can detect fine-scale network structure, such as variations in the probabilities of transitions between items. We next present recent experiments that directly control for differences in transition probabilities, demonstrating that human behavior depends critically on the mesoscale and macroscale properties of networks. Finally, we introduce computational models of human graph learning that make testable predictions about the impact of network structure on people's behavior and cognition. Throughout, we highlight open questions in the study of graph learning that will require creative insights from cognitive scientists and network scientists alike.
\end{abstract}

\dates{This manuscript was compiled on \today}
\doi{\url{www.pnas.org/cgi/doi/10.1073/pnas.XXXXXXXXXX}}

\begin{document}

\maketitle
\thispagestyle{firststyle}
\ifthenelse{\boolean{shortarticle}}{\ifthenelse{\boolean{singlecolumn}}{\abscontentformatted}{\abscontent}}{}

% If your first paragraph (i.e. with the \dropcap) contains a list environment (quote, quotation, theorem, definition, enumerate, itemize...), the line after the list may have some extra indentation. If this is the case, add \parshape=0 to the end of the list environment.
%\dropcap{P}ut the introduction here without titling the intro section.

Our experience of the world is punctuated by discrete items and events, all connected by a hidden network of forces, causes, and associations. Just as navigation requires a mental map of one's physical surroundings \cite{Tolman-01, Golledge-01}, anticipation, planning, perception, and communication all depend on a person's ability to learn the network structure connecting items and events in their environment \cite{Kosko-01, Portugali-01, Baronchelli-01}. For example, in order to identify the boundaries between words, children as young as eight months old identify subtle variations in the network of transitions between syllables in spoken language \cite{Saffran-01}. Within their first 30 months, toddlers already learn enough words to form complex language networks that exhibit robust structural features \cite{Hills-01, Engelthaler-01, Sizemore-01}. By the time we reach adulthood, graph learning enables us to understand and produce language \cite{Saffran-01, Friederici-01}, flexibly and adaptively learn words \cite{Kachergis-01, Kachergis-02}, parse continuous streams of stimuli \cite{Saffran-01}, build social intuitions \cite{Tompson-01}, perform abstract reasoning \cite{Bousfield-01}, and categorize visual patterns \cite{Fiser-01}. In this way, our ability to learn the structures of networks supports a wide range of cognitive functions.

Our capacity to infer and represent complex relationships has also enabled humans to construct an impressive array of networked systems, from language \cite{Steyvers-01, Dorogovtsev-01, Cancho-01} and music \cite{Liu-02} to social networks \cite{Barabasi-02, Girvan-01}, the Internet \cite{Eriksen-01, Albert-03}, and the web of concepts that constitute the arts and sciences \cite{Newman-02, Martin-01}. Moreover, individual differences in cognition, such as those driven by learning disabilities and age, give rise to variations in the types of network structures that people are able to construct \cite{Beckage-01, Dubossarsky-01}. Therefore, studying how humans learn and represent networks will not only inform our understanding of how we perform many of our basic cognitive functions, but will also shed light on the structure and function of networks in the world around us.

Here, we provide a brief introduction to the field of graph learning, spanning the experimental techniques and network-based models, theories, and intuitions recently developed to study the effects of network structure on human cognition and behavior. Given the highly interdisciplinary nature of the field -- which draws upon experimental methods from cognitive science and linguistics and builds upon computational techniques from network science, information theory, and statistical learning -- we aim to present an accessible overview with simple motivating examples.

We focus particular attention on understanding how people uncover the structure of connections between items in a sequence, such as syllables and words in spoken and written language, concepts in books and classroom lectures, or notes in musical progressions. We begin by discussing experimental results demonstrating that humans are adept at detecting differences in the probabilities of transitions between items, and how such transitions connect and combine to form networks that encode the large-scale structure of entire sequences. We then present recent experiments that measure the effects of network structure on human behavior by directly controlling for differences in transition probabilities, followed by a description of the computational models that have been proposed to account for these network effects. We conclude by highlighting some of the open research directions stemming from recent advances in graph learning, including important generalizations of existing graph learning paradigms and direct implications for understanding the structure and function of real-world networks.

\section*{Learning Transition Probabilities}

As humans navigate their environment and accumulate experience, one of the brain's primary functions is to infer the statistical relationships governing causes and effects \cite{Koechlin-01, Wilensky-01}. Given a sequence of items, perhaps the simplest statistics available to a learner are the frequencies of transitions from one item to another. Naturally, the field of statistical learning, which is devoted to understanding how humans extract statistical regularities from their environment, has predominantly focused on these simple statistics. For example, consider spoken language, wherein distinct syllables transition from one to another in a continuous stream without pauses or demarcations between words \cite{Brent-01}. How do people segment such continuous streams of data, identifying where one word starts and another begins? The answer, as research has robustly established \cite{Romberg-01, Aslin-01, Aslin-02, Schapiro-02}, lies in the statistical properties of the transitions between syllables.

\begin{figure}[t]
\centering
\includegraphics[width=.79\linewidth]{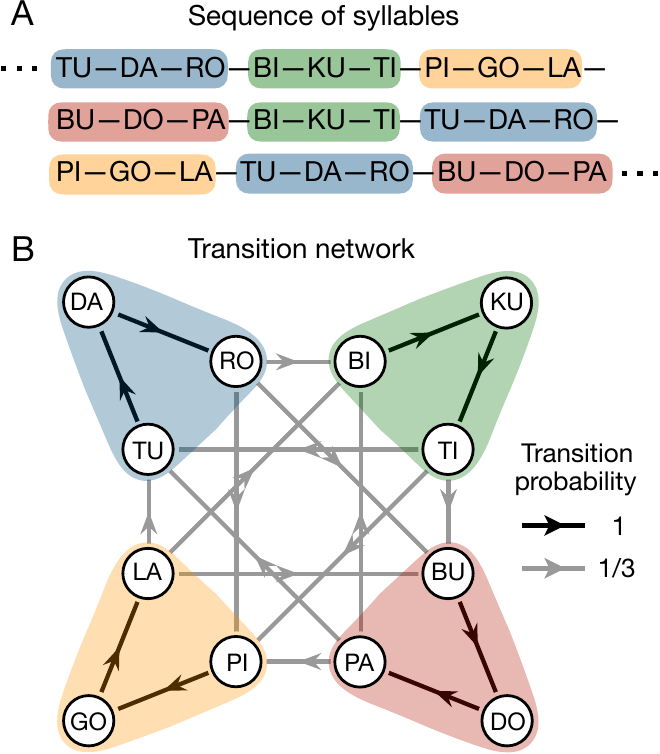}
\caption{Transitions between syllables in the fabricated language of Saffran et al. \cite{Saffran-01}. (A) A sequence containing four different pseudowords: \textit{tudaro} (blue), \textit{bikuti} (green), \textit{budopa} (red), and \textit{pigola} (yellow). When spoken, the sequence forms a continuous stream of syllables, without clear boundaries between words. The transition probability from one syllable to another is 1 if the transition occurs within a word and 1/3 if the transition occurs between words. This difference in transition probabilities allows infants to segment spoken language into distinct words \cite{Saffran-01, Romberg-01, Karuza-01}. (B) Transitions between syllables form a network, with edge weights representing syllable transition probabilities. A random walk in the transition network defines a sequence of syllables in the pseudolanguage. The four pseudowords form distinct communities (highlighted) that are easily identifiable by eye. Reprinted from \cite{Karuza-01} with permission from Elsevier.}
\label{fig_trans}
\end{figure}

The ability to detect words within continuous speech was initially demonstrated by Saffran et al. \cite{Saffran-01}, who exposed infants to sequences of four pseudowords, each consisting of three syllables (Fig. \ref{fig_trans}A). The order of syllables within each word remained consistent, yielding a within-word transition probability of 1. However, the order of the words was random, yielding a between-word transition probability of 1/3. Infants were able to reliably detect this difference in syllable transition probabilities, thereby providing a compelling mechanism for word identification during language acquisition. This experimental paradigm has since been generalized to study statistical learning in other domains, with stimuli ranging from colors \cite{Turk-02} and shapes \cite{Fiser-01} to visual scenes \cite{Brady-01} and  physical actions \cite{Baldwin-01}. Indeed, the capacity to uncover variations in transition probabilities is now recognized as a central and general feature of human learning \cite{Romberg-01, Aslin-01, Aslin-02, Schapiro-02}.

\section*{Learning Network Structure}

Although individual connections between items provide important information about the structure of a system, they do not tell the whole story. Connections also combine and overlap to form complex webs that characterize the higher-order structure of our environment. To study these structures, scientists have increasingly turned to the language of network science \cite{Newman-01}, conceptualizing items as nodes in a network with edges defining possible connections between them (see Supplementary Fig. 1 for a primer on networks). One can then represent a sequence of items, such as the stream of syllables in spoken language, as a walk through this underlying network \cite{Schapiro-01, Kahn-01, Lynn-06, Liu-02, Garvert-01}. This perspective has been particularly useful in the study of artificial grammar learning \cite{Reber-01, Cleeremans-01, Gomez-02}, wherein human subjects are tasked with inferring the grammar rules (i.e., the network of transitions between letters and words) underlying a fabricated language.

By translating items and connections into the language of network science, one inherits a powerful set of descriptive tools and visualization techniques for characterizing different types of structures. For example, consider once again the statistical learning experiment of Saffran et al. (\citealp{Saffran-01}; Fig. \ref{fig_trans}A). Simply by visualizing the transition structure as a network (Fig. \ref{fig_trans}B), it becomes clear that the syllables split naturally into four distinct clusters corresponding to the four different words in the artificial language. This observation raises an important question: When parsing words (or performing any other learning task), are people only sensitive to differences in individual connections, or do they also uncover large-scale features of the underlying network? In what follows, we describe recent advances in graph learning that shed light on precisely this question.

\subsection*{Learning Local Structure}

The simplest properties of a network are those corresponding to individual nodes and edges, such as the weight of an edge, which determines the strength of the connection between two nodes, and the degree of a node, or its number of connections. For example, edge weights can represent transition probabilities between syllables or words \cite{Romberg-01, Aslin-01, Aslin-02, Schapiro-02}, similarities between different semantic concepts \cite{Steyvers-01, Baronchelli-01}, or strengths of social interactions \cite{Barabasi-02, Girvan-01}. Meanwhile, significant effort has focused on understanding how humans learn the network structure surrounding individual nodes \cite{Engelthaler-01, Adelman-01, Balota-01, Carlson-01, Chan-01, Goldstein-01, Yates-01}. For example, the degree defines the connectedness of a node, such as the number of links pointing to a website \cite{Eriksen-01, Masucci-01, Albert-03}, the number of friends that a person has \cite{Barabasi-02}, or the number of citations accumulated by a scientific paper \cite{Martin-01}. Notably, many of the networks that people encounter on a daily basis -- including language, social, and hyperlink networks -- exhibit heavy-tailed degree distributions, with many nodes of low degree and a select number of high-degree hubs \cite{Steyvers-01, Baronchelli-01, Dorogovtsev-01, Cancho-01, Eriksen-01, Masucci-01, Barabasi-01, Newman-02, Borge-01}.

\begin{figure}[t!]
\centering
\includegraphics[width=\linewidth]{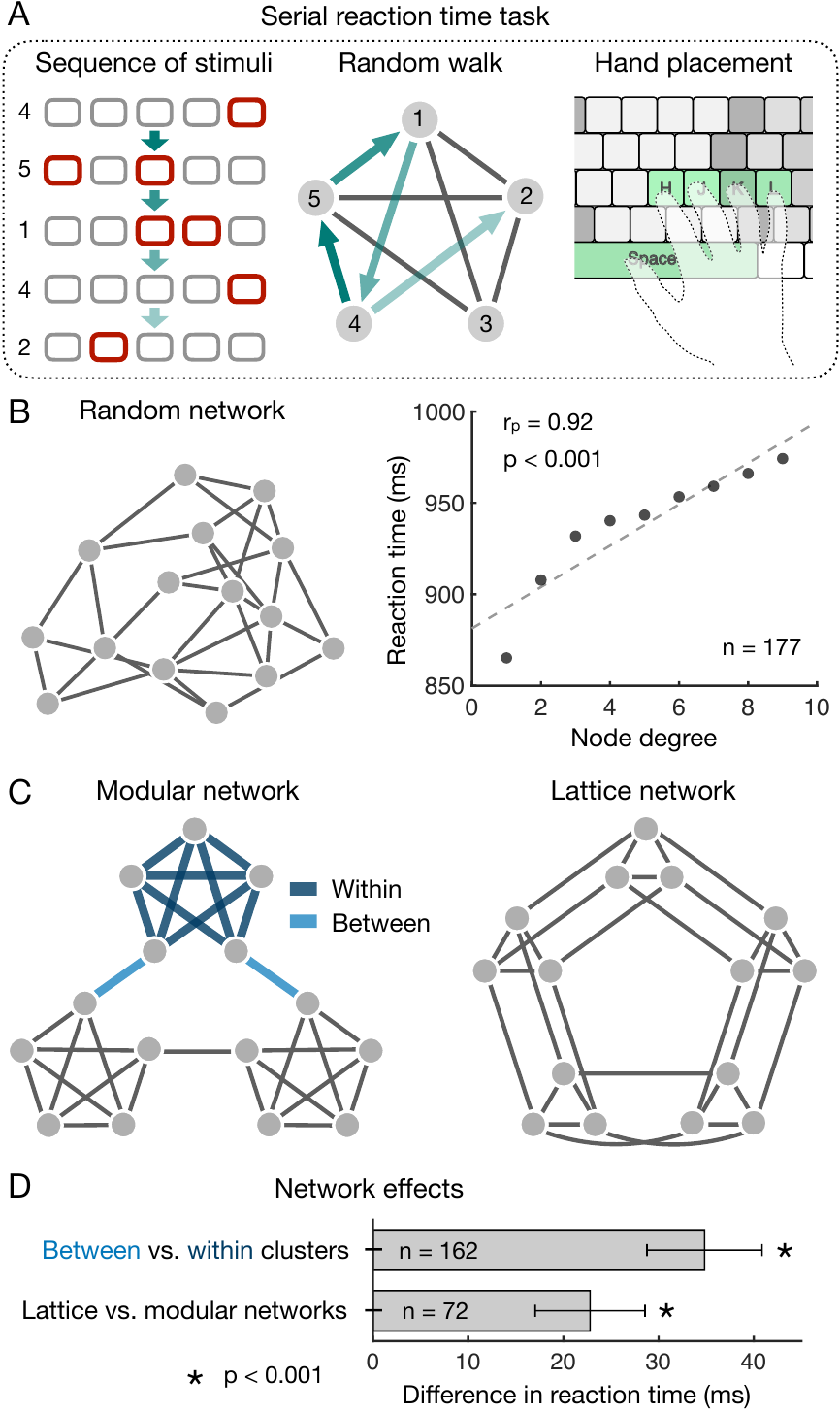}
\caption{Human behavior depends on network topology. (A) We consider a serial reaction time experiment in which subjects are shown sequences of stimuli and are asked to respond by performing an action. Here, each stimulus consists of five squares, one or two of which are highlighted in red (left); the order of stimuli is determined by a random walk on an underlying network (center); and for each stimulus, the subject presses the keys on the keyboard corresponding to the highlighted squares (right). (B) Considering Erd\"{o}s-R\'{e}nyi random transition networks with 15 nodes and 30 edges (left), subjects' average reaction times to a transition $i\rightarrow j$ increase as the degree $k_i$ of the preceding node increases (right). Equivalently, subjects' reaction times increase as the transition probability $P_{ij} = 1/k_i$ decreases \cite{Lynn-06}. (C) To control for variations in transition probabilities, we consider two networks with constant degree $k=4$: a \textit{modular network} consisting of three communities of five nodes each (left) and a \textit{lattice network} representing a 3$\times$5 grid with periodic boundary conditions (right). (D) Experiments indicate two consistent effects of network structure. First, in the modular network, reaction times for between-cluster transitions are longer than for within-cluster transitions \cite{Karuza-02, Kahn-01, Karuza-03, Lynn-06}. Second, reaction times are longer on average for the lattice network than for the modular network \cite{Kahn-01, Lynn-06}.}
\label{fig_effects}
\end{figure}

Significant research has now demonstrated that people are able to learn the local network properties of individual nodes and edges, such as the transition probabilities between syllables in the previous section \cite{Romberg-01, Aslin-01, Aslin-02, Schapiro-02}. To illustrate the impact of network structure on human behavior, we consider a recently-developed experimental paradigm \cite{Kahn-01, Lynn-06}, while noting that similar results have also been achieved using variations on this approach \cite{Schapiro-01, Karuza-01, Garvert-01, Karuza-02, Karuza-03, Tompson-01}. Specifically, each subject is shown a sequence of stimuli, with the order of stimuli defined by a random walk on an underlying transition network (Fig. \ref{fig_effects}A). Subjects are asked to respond to each stimulus by performing an action (and to avoid confounds the assignment of stimuli to nodes in the network is randomized across subjects). By measuring the speed with which subjects respond to stimuli, one can infer their expectations about the network structure: A fast reaction reflects a strongly-anticipated transition, while a slow reaction reflects a weakly-anticipated (or surprising) transition \cite{Hyman-01, McCarthy-01, Kahn-01, Lynn-06}.

Intuitively, one should expect a subject's anticipation to increase (and thus their reaction time to decrease) for edges representing more probable transitions. In order to test this prediction, we note that for a random walk in an unweighted and undirected network, the transition probability from one node $i$ to a neighboring node $j$ is given by $P_{ij} = 1/k_i$, where $k_i$ is the degree of node $i$. Aligning with intuition, researchers have shown that people's reaction times are positively correlated with the degree of the previous stimulus (Fig. \ref{fig_effects}B), and therefore, people are better able to anticipate more probable transitions \cite{Kahn-01, Lynn-06}. Interestingly, significant research has also established similar results in language networks, with people reading words more quickly if they occur more frequently or appear in more contexts \cite{Forster-01, Balota-01, Adelman-01}. Conversely, humans tend to slow down and produce more errors when attempting to recall words with a large number of semantic associations, a phenomenon known as the fan effect \cite{Anderson-01, Anderson-02}. Together, these results demonstrate that humans are sensitive to variations in the local properties of individual nodes and edges, but what about the mesoscale and macroscale properties of a network?

\subsection*{Learning Mesoscale Structure}

The mesoscale structure of a network reflects the organizational properties of groups of nodes and edges. One such property is clustering, or the tendency for a pair of nodes with a common neighbor to form a connection themselves. This tendency is clearly observed in social networks, where people with a common friend are themselves more likely to become friends. Similar principles govern the mesoscale structure of many other real-world networks, with items such as words, scientific papers, and webpages all exhibiting high clustering \cite{Motter-01, Sigman-01, Watts-01, Martin-01}. As nodes cluster together, they often give rise to a second mesoscale property -- modular structure -- which is characterized by tightly-connected modules or communities of nodes. Such modular structure is now recognized as a ubiquitous feature of networks in our environment \cite{Newman-03}, with language splitting into groups of semantically or phonetically similar words \cite{Bousfield-01, Cancho-01}, people forming social cliques \cite{Moody-01, Barabasi-02, Girvan-01}, and websites clustering into online communities \cite{Eriksen-01}.

Over the past ten years, researchers have made signifiant strides toward understanding how the mesoscale properties of a network impact human learning and behavior. Words with higher clustering are more likely to be acquired during language learning \cite{Goldstein-01}, while words with lower clustering are easier to recognize in long-term memory \cite{Vitevitch-01} and convey processing \cite{Chan-01, Yates-01} and production \cite{Chan-02} benefits. Additionally, in a series of cognitive and neuroimaging experiments, researchers have found that a network's modular structure has a significant impact on human behavior and neural activity. For example, people are able to detect the boundaries between communities in a network just by observing sequences of nodes \cite{Schapiro-01, Karuza-02, Kahn-01, Karuza-03, Lynn-06}. Moreover, strong modular structure helps people build more accurate mental representations of a network, thereby allowing humans to better anticipate future items and events \cite{Schapiro-01, Karuza-02, Kahn-01, Karuza-03, Lynn-06}.

\subsection*{Learning Global Structure}

In addition to their local and mesoscale features, networks also have global properties that depend on the entire architecture of nodes and edges. Perhaps the most well-studied global property is small-world structure, wherein each node connects to every other node in only a small number of steps \cite{Watts-01}. Small-world topology has been observed in an array of networks that humans are tasked with learning, including social relationships \cite{Kleinberg-01}, web hyperlinks \cite{Albert-03}, scientific citations \cite{Martin-01}, and semantic associations in language \cite{Borge-01, Cancho-01}. Moreover, in a particularly compelling example of the relationship between global network structure and human cognition, the small-world structure of people's learned language networks has been shown to vary from person to person, decreasing with age \cite{Dubossarsky-01} and in people with learning disabilities \cite{Beckage-01}.

While small-worldness describes the structure of an entire network, there are also measures that relate individual nodes to a network's global topology, including centrality (a measure of a node's role in mediating long-distance connections), communicability (a measure of the number of paths connecting a pair of nodes), and coreness (a measure of how deeply embedded a node is in a network). Global measures such as these have recently been shown to impact human learning and cognition, indicating that humans are sensitive to the global structure of networks in their environment. For example, in the reaction time experiments described above (Fig. \ref{fig_effects}A), people responded more quickly, and therefore were better able to anticipate, nodes with low centrality \cite{Kahn-01}. In a related experiment, neural activity was shown to reflect the communicability between pairs of stimuli in an underlying transition network \cite{Garvert-01}. Finally, as children learn language, they more readily acquire and produce words with low coreness \cite{Carlson-01}. Together, these results point to a robust and general relationship between large-scale network structure and human cognition. However, might these large-scale network effects simply be driven by confounding variations in the local network structure?

\subsection*{Controlling for Differences in Local Structure}

To disentangle the effects of large-scale network structure from those of local structure, recent research has directly controlled for differences in transition probabilities by focusing on specific families of networks \cite{Schapiro-01, Karuza-02, Kahn-01, Lynn-06}. Recall that for random walks on unweighted, undirected networks, the transition probabilities are determined by node degrees. Therefore, to ensure that all transitions have equal probability, one can simply focus on graphs with constant degree but varying topology. For example, consider the modular and lattice graphs shown in Fig. \ref{fig_effects}C. Since both networks have constant degree 4 (and therefore constant transition probability 1/4 across all edges), any variation in behavior or cognition between different parts of a network, or between the two networks themselves, must stem from the networks' global topologies.

This approach was first developed by Schapiro et al. \cite{Schapiro-01}, who demonstrated that people are able to detect the transitions between clusters in the modular graph (Fig. \ref{fig_effects}C), and that these between-cluster transitions yield distinct patterns of neural activity relative to within-cluster transitions. Returning to the reaction time experiment (Fig. \ref{fig_effects}A), it was shown that subjects react more quickly to (and therefore are able to better anticipate) within-cluster transitions than between-cluster transitions (\citealp{Kahn-01, Lynn-06}; Fig \ref{fig_effects}D). Moreover, people exhibit an overall decrease in reaction times for the modular graph relative to the lattice graph (\citealp{Kahn-01, Lynn-06}; Fig. \ref{fig_effects}D).

These results, combined with findings in similar experiments \cite{Karuza-02, Karuza-03}, demonstrate that humans are sensitive to features of mesoscale and global network topology, even after controlling for differences in local structure. Thus, not only are humans able to learn individual transition probabilities, as originally demonstrated in seminal statistical learning experiments (Fig. \ref{fig_trans}), they are also capable of uncovering some of the complex structures found in our environment. But how do people learn the large-scale features of networks from past observations?

\section*{Modeling Human Graph Learning}

Experiments spanning cognitive science, neuroscience, linguistics, and statistical learning have established that human behavior and cognition depend on the mesoscale and global topologies of networks in their environment. To understand how people detect these global features, and to make quantitative predictions about human behavior, one requires computational models of how humans construct internal representations of networks from past experiences. Here, we again focus on understanding how people learn the networks of transitions underlying observed sequences of items, such as words in a sentence, concepts in a book or classroom lecture, or notes in a musical progression. Interestingly, humans systematically deviate from the most accurate, and perhaps the simplest, learning rule.

To make these ideas concrete, consider a sequence of items described by the transition probability matrix $P$, where $P_{ij}$ represents the conditional probability of one item $i$ transitioning to another item $j$. Given an observed sequence of items, one can imagine estimating $P_{ij}$ by simply dividing the number of times $i$ has transitioned to $j$ (denoted by $n_{ij}$) by the number of times $i$ has appeared (which equals $\sum_k n_{ik}$):
\begin{equation}
\label{MLE}
\hat{P}_{ij} = \frac{n_{ij}}{\sum_k n_{ik}}.
\end{equation}
In fact, not only is this perhaps the simplest estimate one could perform, it is also the most accurate (or maximum likelihood) estimate of the transition probabilities from past observations \cite{Boas-01}. An important feature of maximum likelihood estimation is that it gives an unbiased approximation of the true transition probabilities; that is, the estimated transition probabilities $\hat{P}_{ij}$ are evenly distributed about their true values $P_{ij}$, independent of the large-scale structure of the network \cite{Boas-01}. However, we have seen that people's behavior and cognition depend systematically on mesoscale and global network properties, even when transition probabilities are held constant \cite{Schapiro-01, Garvert-01, Karuza-02, Kahn-01, Lynn-06}. Thus, when constructing internal representations, humans allow higher-order network structure to influence their estimates of individual transition probabilities, thereby deviating from maximum likelihood estimation \cite{Lynn-06}.

To understand the impact of network topology on human cognition, researchers have recently proposed a number of models describing how humans learn and represent transition networks \cite{Howard-01, Dehaene-01, Meyniel-01, Angela-01, Lynn-06, Garvert-01, Meyniel-02, Momennejad-01}. Notably, many of these models share a common underlying mechanism: that instead of just counting transitions of length one (as in maximum likelihood estimation), humans also include transitions of lengths two, three, or more in their representations \cite{Angela-01, Lynn-06, Garvert-01, Meyniel-02, Momennejad-01, Newport-01}. Mathematically, by combining transitions of different distances, the estimated transition probabilities take the form
\begin{equation}
\hat{P}_{ij} = C \sum_{t \ge 1} f(t) n_{ij}^{(t)},
\end{equation}
where $n_{ij}^{(t)}$ represents the number of times that $i$ has transitioned to $j$ in $t$ steps, $f(t)$ defines the weight placed on transitions of a given distance, and $C$ is a normalization constant. Interestingly, this simple prediction can be derived from a number of different cognitive theories -- including the temporal context model of episodic memory \cite{Howard-01}, temporal difference learning and the successor representation in reinforcement learning \cite{Sutton-02, Dayan-01, Gershman-01}, and the free energy principle from information theory \cite{Lynn-06}. But how does combining transitions over different distances allow people to learn the structure of a network?

%\begin{figure*}[t]
\begin{figure}[t!]
\centering
\includegraphics[width=.8\linewidth]{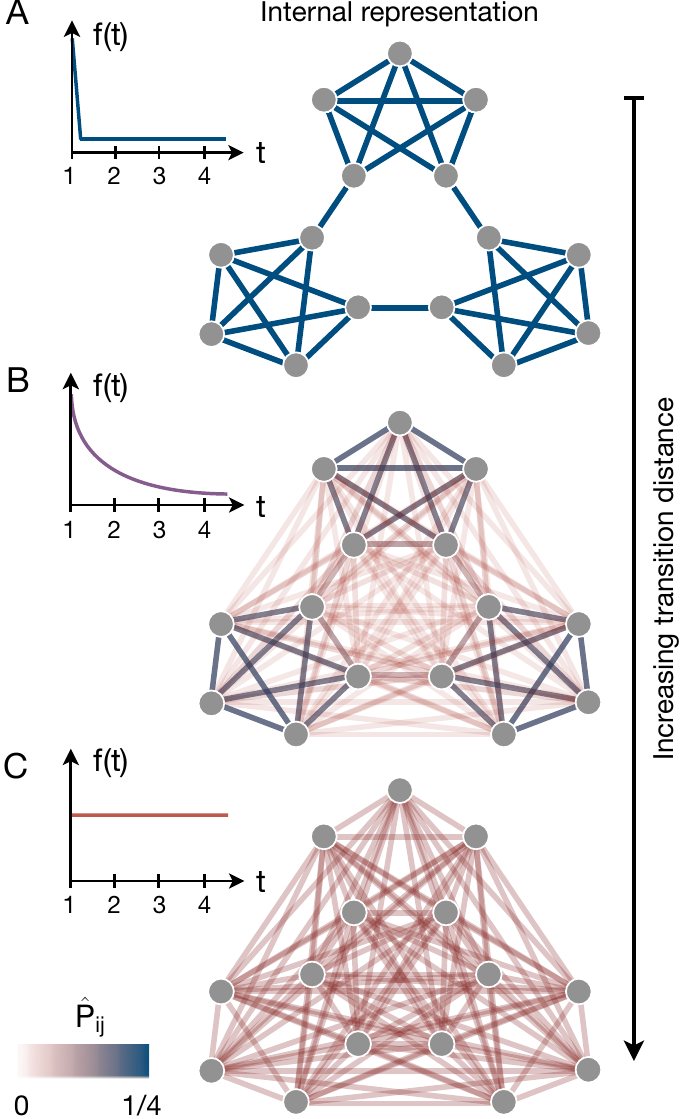}
\caption{Mesoscale and global network features emerge from long-distance associations. (A) Illustration of the weight function $f(t)$ (left) and the learned network representation $\hat{P}$ for learners that only consider transitions of length one. The estimated structure resembles the true modular network. (B) For learners that down-weight transitions of longer distances, higher-order features of the transition network, such as community structure, organically come into focus, yielding higher expected probabilities for within-cluster transitions than for between-cluster transitions. (C) For learners that equally weigh transitions of all distances, the internal representation becomes all-to-all, losing any resemblance to the true transition network. Panels A through C correspond to learners that include progressively longer transitions in their network estimates. Adapted from \cite{Lynn-06}.}
\label{fig_models}
%\end{figure*}
\end{figure}

To answer this question, it helps to consider different choices for the function $f(t)$. Typically, $f(t)$ is assumed to be decreasing such that longer-distance associations contribute more weakly to a person's network representation \cite{Sutton-02, Gershman-01, Lynn-06}. If $f(t)$ is a delta function centered at $t = 1$ (Fig. \ref{fig_models}A), then the learner focuses on transitions of length one. In this case, people simply perform maximum likelihood estimation, resulting in an unbiased estimate of the true transition structure $P$. Conversely, if $f(t)$ is uniform over all time scales $t\ge 1$, then the learner equally weighs transitions of all distances (Fig. \ref{fig_models}C), and the estimate $\hat{P}$ loses any resemblance to the true transition structure $P$. Importantly, however, for learners who combine transitions over intermediate distances (Fig. \ref{fig_models}B), we find that large-scale features of the network organically come into focus. Consider, for example, the modular network from Fig. \ref{fig_effects}C. By combining transitions of lengths two, three, or more, humans tend to over-weigh the associations within communities and under-weigh the transitions between communities (Fig. \ref{fig_models}B). This simple observation explains why people are surprised by cross-cluster transitions (\citealp{Kahn-01, Lynn-06}; Fig. \ref{fig_effects}D), why sequences in lattice and random networks are more difficult to anticipate (\citealp{Kahn-01, Lynn-06}; Fig. \ref{fig_effects}D), and how people detect the boundaries between clusters \cite{Schapiro-01, Karuza-02, Karuza-03}.

More generally, the capacity to learn the large-scale structure of a network enables people to perform many basic cognitive functions, from anticipating non-adjacent dependencies between syllables and words \cite{Newport-01, Altmann-01} to planning for future events \cite{Atance-01, Addis-01} and estimating future rewards \cite{Sutton-02, Gershman-01}. Using models similar to that above, researchers have been able to predict the impacts of network structure on human behavior in reinforcement learning tasks \cite{Momennejad-01}, pattern detection in random sequences \cite{Angela-01, Meyniel-02}, and variations in neural activity \cite{Schapiro-01, Garvert-01, Meyniel-02}. Notably, the explained effects span various types of behavioral and neural observations, including reaction times \cite{Huettel-01, Kahn-01, Lynn-06}, data segmentation \cite{Schapiro-01, Karuza-02, Karuza-03}, task errors \cite{Kahn-01, Lynn-06}, randomness detection \cite{Falk-01}, EEG signals \cite{Squires-01}, and fMRI recordings \cite{Huettel-01, Schapiro-01}. Together, these results indicate that people's ability to detect the mesoscale and global structure of a network emerges not just from their capacity to learn individual edges, but also from their capacity to associate items across spatial, temporal, and topological scales.

\section*{The Future of Graph Learning}

Past and current advances in graph learning inspire new research questions at the intersection of cognitive science, neuroscience, and network science. Here, we highlight a number of important directions, beginning with possible generalizations of the existing graph learning paradigm before discussing the implications of graph learning for our understanding of the structures and functions of real-world transition networks.

\subsection*{Extending the Graph Learning Paradigm}

Most graph learning experiments, including those discussed in Figs. \ref{fig_trans} and \ref{fig_effects}, present each subject with a sequence of stimuli defined by a random walk on a (possibly weighted and directed) transition network \cite{Saffran-01, Schapiro-01, Karuza-01, Karuza-02, Kahn-01, Newport-01, Garvert-01, Reber-01, Cleeremans-01, Gomez-02, Karuza-03, Tompson-01, Lynn-06}. Equivalently, in the language of stochastic processes, each sequence represents a stationary Markov process \cite{Ross-01}. Although random walks offer a natural starting point in the study of graph learning, they are also constrained by three main assumptions: (i) that the underlying transition structure remains static over time (stationarity), (ii) that future stimuli only depend on the current stimulus (the Markov property), and (iii) that the sequence is predetermined without input from the observer. Future graph learning experiments can test the boundaries of these constraints by systematically generalizing the existing graph learning paradigm.

\subsubsection*{Stationarity}

\begin{figure}[t]
\centering
\includegraphics[width=\linewidth]{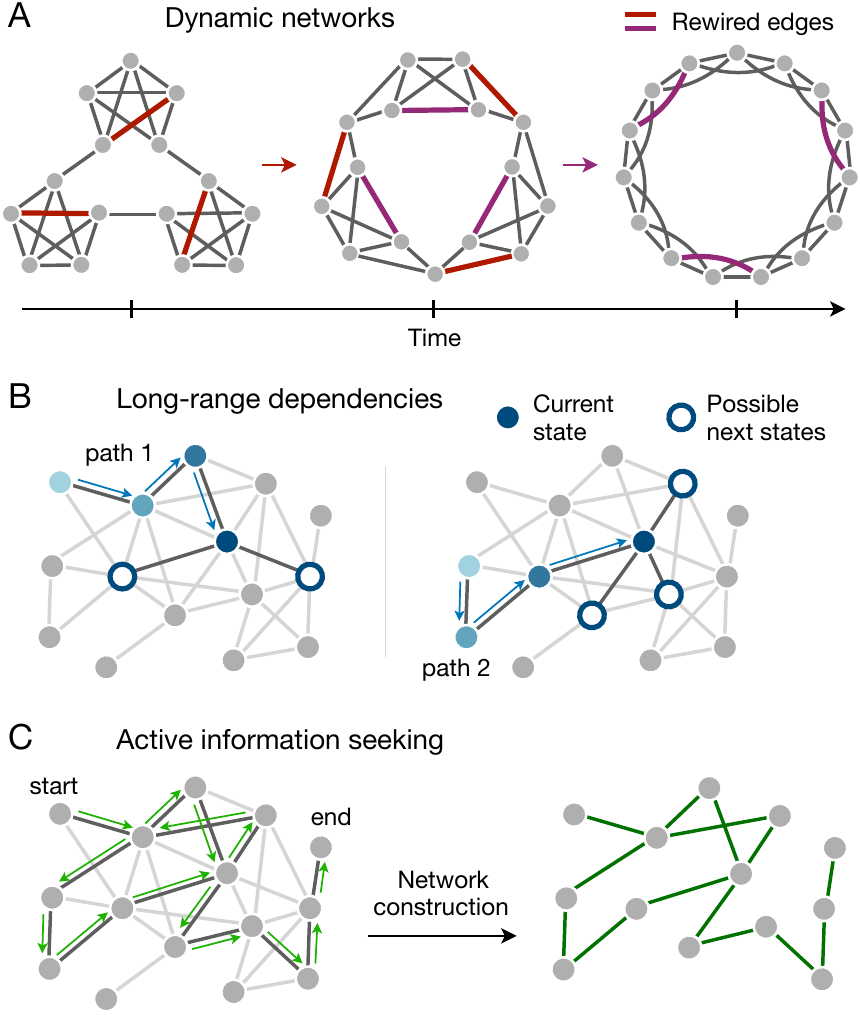}
\caption{Generalizations of the graph learning paradigm. (A) Transition networks often shift and change over time. Such non-stationary transition probabilities can be described using dynamical transition networks, which evolve from one network (for example, the modular network on the left) to another (for example, the ring network on the right) by iteratively rewiring edges. (B) Many real-world sequences have long-range dependencies, such that the next state depends not just on the current state, but also on a number of previous states \cite{Amit-01, Jafari-01}. For example, path 1 in the displayed network yields two possibilities for the next state (left), while path 2 yields a different set of three possible states (right). (C) Humans often actively seek out information by choosing their path through a transition network, rather than simply being presented with a prescribed sequence. Such information seeking yields a subnetwork containing the nodes and edges traversed by the walker.}
\label{fig_general}
\end{figure}

While most graph learning experiments focus on static transition networks, many of the networks that humans encounter in the real world either evolve in time or overlap with other networks in the environment \cite{Dorogovtsev-01, Sizemore-01, Steyvers-01, Beckage-01, Barabasi-02}. Therefore, rather than simply investigating people's ability to learn a single network, future experiments should study the capacity for humans to detect the dynamical features of an evolving network (Fig. \ref{fig_general}A) or differentiate the distinct features of multiple networks. Early results indicate that, when observing a sequence of stimuli that shifts from one transition structure to another, people's learned representation of the first network influences their behavior in response to the second network, but that these effects diminish with time \cite{Kahn-01}. This gradual ``unlearning" of network structure raises an important question for future research: Rather than investigating how network properties facilitate learning -- as has been the focus of most graph learning studies -- can we determine which properties make a network difficult to forget?

%\begin{figure}[b!]

\subsubsection*{The Markov Property}

Thus far, in keeping with the majority of existing graph learning research, we have focused exclusively on sequences in which the next stimulus depends only on the current stimulus; that is, we have focused on sequences that obey the Markov property \cite{Ross-01}. However, almost all sequences of stimuli or items in the real world involve long-range correlations and dependencies (Fig. \ref{fig_general}B). For example, the probability of a word in spoken language depends not just on the previous word, but also the earlier words in the sentence and the broader context in which the sentence exists \cite{Amit-01}. Similarly, musical systems often enforce constraints on the length and structure of sequences, thereby inducing long-range dependencies between notes \cite{Jafari-01}. Interestingly, given mounting evidence that people construct long-distance associations \cite{Angela-01, Lynn-06, Garvert-01, Meyniel-02, Momennejad-01, Newport-01}, the resulting internal estimates of transition structures resemble non-Markov processes \cite{Lynn-06}. Therefore, future research could investigate whether the learning long-distance associations enables people to infer the non-Markov features of sequences in daily life.

%\begin{figure*}[t]
%\centering
%\includegraphics[width=.8\textwidth]{Fig6_learnability_2.pdf}
%\caption{Real transition networks exhibit hierarchical structure. (A) A language network constructed from the words (nodes) and transitions between them (edges) in the complete works of Shakespeare. (B) A music network representing the transitions between notes in Beethoven's Sonata No. 23. (C) A knowledge network defined by hyperlinks between pages on Wikipedia. (D, E) Many real-world transition networks exhibit hierarchical organization \cite{Ravasz-01}, which is characterized by two topological features: (D) Heterogeneous structure, which is often associated with scale-free networks, is typically characterized by a power-law degree distribution and the presence of high-degree hub nodes \cite{Barabasi-01}. (E) Modular structure is defined by the presence of clusters of nodes with dense within-cluster connectivity and sparse between-cluster connectivity \cite{Girvan-01}.}
%\label{fig_learnability}
%\end{figure*}

\subsubsection*{Information Seeking}

Finally, although many of the sequences that humans observe are prescribed without input from the observer, there are also settings in which people have agency in determining the structure of a sequence. For example, when surfing the Internet \cite{Adamic-02, Dodds-01, Oday-01, West-01} or following a trail of scientific citations \cite{Martin-01}, people choose their paths through the underlying hyperlink and citation networks. In this way, people are able to seek out information about networks structures rather than simply having the information presented to them (Fig. \ref{fig_general}C). Such information seeking has been shown to vary by person \cite{Oday-01} and to depend crucially on the topology of the underlying network \cite{Adamic-02, Dodds-01, West-01}. Moreover, when retrieving information from memory, humans search through their stored networks of associations \cite{Raaijmakers-02}, often performing search strategies that resemble optimal foraging in physical space \cite{Hills-02, Jones-01, Avery-01}. In the context of graph learning, allowing subjects to actively seek information raises a number of compelling questions: Does choosing their path through a transition network enable subjects to more efficiently learn its topology? Or does the ability to seek information lead people to form biased representations of the true transition structure \cite{Schulz-01, Jonas-01}? These questions, combined with the directions described above, highlight some of the exciting extensions of graph learning that will require creative insights and collaborative contributions from cognitive scientists and network scientists alike.

\subsection*{Studying the Structure of Real-World Networks}

\begin{figure}[t]
\centering
\includegraphics[width=\linewidth]{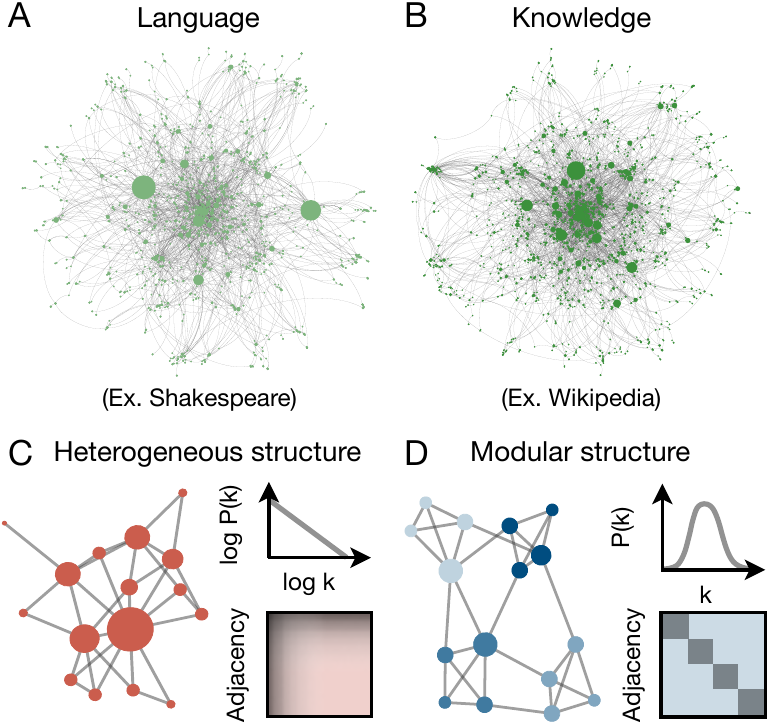}
\caption{Real transition networks exhibit hierarchical structure. (A) A language network constructed from the words (nodes) and transitions between them (edges) in the complete works of Shakespeare. (B) A knowledge network of hyperlinks between pages on Wikipedia. (C, D) Many real-world transition networks exhibit hierarchical organization \cite{Ravasz-01}, which is characterized by two topological features: (C) Heterogeneous structure, which is often associated with scale-free networks, is typically characterized by a power-law degree distribution and the presence of high-degree hub nodes \cite{Barabasi-01}. (D) Modular structure is defined by the presence of clusters of nodes with dense within-cluster connectivity and sparse between-cluster connectivity \cite{Girvan-01}.}
\label{fig_learnability}
\end{figure}

In addition to shedding light on human behavior and cognition, the study of graph learning also has the promise to offer insights into the structure and function of real-world networks. Indeed, there exists an intimate connection between human cognition and networks: While people rely on networked systems to perform a wide range of tasks, from communicating using language (Fig. \ref{fig_learnability}A) and music (Fig. \ref{fig_learnability}B) to storing and retrieving information through science and the Internet (Fig. \ref{fig_learnability}C), many of these networks have evolved with or were explicitly designed by humans. Therefore, just as humans are adept at learning the structure of networks, one might suspect that some networks are structured to support human learning and cognition.

The perspective that cognition may constrain network structure has recently shed light on the organizational properties of some real-world networks \cite{Baronchelli-01, Borge-01}, including the small-world structure and power-law degree distributions exhibited by semantic and word co-occurrence networks \cite{Steyvers-01, Dorogovtsev-01, Cancho-01}, and the scale-free structure of the connections between concepts on Wikipedia \cite{Masucci-01}. Interestingly, many of the networks with which humans interact share two distinct structural features: (i) They are heterogeneous (Fig. \ref{fig_learnability}D), characterized by the presence of hub nodes with unusually high degree \cite{Barabasi-01, Cancho-01, Newman-02, Steyvers-01, Borge-01}, and (ii) they are modular (Fig. \ref{fig_learnability}E), characterized by the existence of tightly-connected clusters \cite{Girvan-01, Motter-01, Eriksen-01, Steyvers-01, Borge-01}. Together, heterogeneity and modularity represent the two defining features of hierarchical organization, which has now been observed in a wide array of man-made networks \cite{Ravasz-01, Arenas-02}. Could it be that the shared structural properties of these networks arise from their common functional purpose: to facilitate human learning and communication?

Graph learning provides quantitative models and experimental tools to begin answering questions such as these. For example, experimental results, such as those discussed in Fig. \ref{fig_effects}, indicate that modular structure improves people's ability to anticipate transitions \cite{Kahn-01, Lynn-06}, and this result has been confirmed numerically using models of the form in Fig. \ref{fig_models} \cite{Lynn-06}. Moreover, the high-degree hubs found in heterogeneous networks have been shown to help people search for information \cite{Adamic-02, West-01}. Together, these results demonstrate that graph learning offers a unique and constructive lens through which to study networks in the world around us.

\section*{Conclusions and Outlook}

Understanding how people learn and represent the complex relationships governing their environment remains one of the greatest open problems in the study of human cognition. On the heels of decades of research in cognitive science and statistical learning investigating how humans detect the local properties of individual items and the connections between them \cite{Saffran-01, Turk-02, Fiser-01, Brady-01, Baldwin-01, Romberg-01, Aslin-01, Aslin-02, Schapiro-02}, conclusive evidence now demonstrates that human behavior, cognition, and neural activity depend critically on the large-scale structure of items and connections \cite{Schapiro-01, Karuza-01, Garvert-01, Karuza-02, Karuza-03, Tompson-01, Kahn-01, Lynn-06}. By casting the items and connections in our environment as nodes and edges in a network, scientists can now explore the impact of network structure on human cognition in a unified and principled framework.

Although the experimental and numerical foundation of the field has been laid, graph learning remains a budding area of research offering a wealth of interdisciplinary opportunities. From new cognitive modeling techniques (Fig. \ref{fig_models}) and extensions of existing experimental paradigms (Fig. \ref{fig_general}) to novel applications in the study of real-world networks (Fig. \ref{fig_learnability}), graph learning is primed to alter the way we think about human cognition, complex networks, and the myriad ways in which they intersect.

\matmethods{The materials and methods discussed in this article are presented and described in the references listed herein.

\section*{Data Availability}

This article contains no new data.
}

\showmatmethods{} % Display the Materials and Methods section

\acknow{We thank David Lydon-Staley, Nico Christianson, and Jennifer Stiso for comments on previous versions of the paper. D.S.B. and C.W.L. acknowledge support from the John D. and Catherine T. MacArthur Foundation, the Alfred P. Sloan Foundation, the ISI Foundation, the Paul G. Allen Family Foundation, the Army Research Laboratory (W911NF-10-2-0022), the Army Research Office (Bassett-W911NF-14-1-0679, Grafton-W911NF-16-1-0474, DCIST- W911NF-17-2-0181), the Office of Naval Research, the National Institute of Mental Health (2-R01-DC-009209-11, R01-MH112847, R01-MH107235, R21-M MH-106799), the National Institute of Child Health and Human Development (1R01HD086888-01), National Institute of Neurological Disorders and Stroke (R01 NS099348), and the National Science Foundation (BCS-1441502, BCS-1430087, NSF PHY-1554488 and BCS-1631550). The content is solely the responsibility of the authors and does not necessarily represent the official views of any of the funding agencies.}

\showacknow{} % Display the acknowledgments section

% Bibliography
\bibliography{GraphLearningBib}

\begin{thebibliography}{102}
\providecommand{\natexlab}[1]{#1}
\providecommand{\url}[1]{\texttt{#1}}
\expandafter\ifx\csname urlstyle\endcsname\relax
  \providecommand{\doi}[1]{doi: #1}\else
  \providecommand{\doi}{doi: \begingroup \urlstyle{rm}\Url}\fi

\bibitem[Tolman(1948)]{Tolman-01}
Edward~C Tolman.
\newblock Cognitive maps in rats and men.
\newblock \emph{Psychol. Rev.}, 55\penalty0 (4):\penalty0 189, 1948.

\bibitem[Golledge(2003)]{Golledge-01}
Reginald~G Golledge.
\newblock Human wayfinding and cognitive maps.
\newblock In \emph{The Colonization of Unfamiliar Landscapes}, pages 49--54.
  Routledge, 2003.

\bibitem[Kosko(1986)]{Kosko-01}
Bart Kosko.
\newblock Fuzzy cognitive maps.
\newblock \emph{Int. J. Man Mach. Stud.}, 24\penalty0 (1):\penalty0 65--75,
  1986.

\bibitem[Portugali(1996)]{Portugali-01}
Juval Portugali.
\newblock \emph{The construction of cognitive maps}, volume~32.
\newblock Springer Science \& Business Media, 1996.

\bibitem[Baronchelli et~al.(2013)Baronchelli, Ferrer-i Cancho, Pastor-Satorras,
  Chater, and Christiansen]{Baronchelli-01}
Andrea Baronchelli, Ramon Ferrer-i Cancho, Romualdo Pastor-Satorras, Nick
  Chater, and Morten~H Christiansen.
\newblock Networks in cognitive science.
\newblock \emph{Trends Cogn. Sci.}, 17\penalty0 (7):\penalty0 348--360, 2013.

\bibitem[Saffran et~al.(1996)Saffran, Aslin, and Newport]{Saffran-01}
Jenny~R Saffran, Richard~N Aslin, and Elissa~L Newport.
\newblock Statistical learning by 8-month-old infants.
\newblock \emph{Science}, 274\penalty0 (5294):\penalty0 1926--1928, 1996.

\bibitem[Hills et~al.(2009)Hills, Maouene, Maouene, Sheya, and Smith]{Hills-01}
Thomas~T Hills, Mounir Maouene, Josita Maouene, Adam Sheya, and Linda Smith.
\newblock Longitudinal analysis of early semantic networks: Preferential
  attachment or preferential acquisition?
\newblock \emph{Psychol. Sci.}, 20\penalty0 (6):\penalty0 729--739, 2009.

\bibitem[Engelthaler and Hills(2017)]{Engelthaler-01}
Tomas Engelthaler and Thomas~T Hills.
\newblock Feature biases in early word learning: Network distinctiveness
  predicts age of acquisition.
\newblock \emph{Cogn. Sci.}, 41:\penalty0 120--140, 2017.

\bibitem[Sizemore et~al.(2018)Sizemore, Karuza, Giusti, and
  Bassett]{Sizemore-01}
Ann~E Sizemore, Elisabeth~A Karuza, Chad Giusti, and Danielle~S Bassett.
\newblock Knowledge gaps in the early growth of semantic feature networks.
\newblock \emph{Nat. Hum. Behav.}, 2\penalty0 (9):\penalty0 682, 2018.

\bibitem[Friederici(2005)]{Friederici-01}
Angela~D Friederici.
\newblock Neurophysiological markers of early language acquisition: {F}rom
  syllables to sentences.
\newblock \emph{Trends Cogn. Sci.}, 9\penalty0 (10):\penalty0 481--488, 2005.

\bibitem[Kachergis et~al.(2012)Kachergis, Yu, and Shiffrin]{Kachergis-01}
George Kachergis, Chen Yu, and Richard~M Shiffrin.
\newblock An associative model of adaptive inference for learning
  word--referent mappings.
\newblock \emph{Psychon. Bull. Rev.}, 19\penalty0 (2):\penalty0 317--324, 2012.

\bibitem[Kachergis et~al.(2013)Kachergis, Yu, and Shiffrin]{Kachergis-02}
George Kachergis, Chen Yu, and Richard~M Shiffrin.
\newblock Actively learning object names across ambiguous situations.
\newblock \emph{Top. Cogn. Sci.}, 5\penalty0 (1):\penalty0 200--213, 2013.

\bibitem[Tompson et~al.(2019)Tompson, Kahn, Falk, Vettel, and
  Bassett]{Tompson-01}
Steven~H Tompson, Ari~E Kahn, Emily~B Falk, Jean~M Vettel, and Danielle~S
  Bassett.
\newblock Individual differences in learning social and nonsocial network
  structures.
\newblock \emph{J. Exp. Psychol. Learn. Mem. Cogn.}, 45\penalty0 (2):\penalty0
  253, 2019.

\bibitem[Bousfield(1953)]{Bousfield-01}
Weston~A Bousfield.
\newblock The occurrence of clustering in the recall of randomly arranged
  associates.
\newblock \emph{J. Gen. Psychol.}, 49\penalty0 (2):\penalty0 229--240, 1953.

\bibitem[Fiser and Aslin(2002)]{Fiser-01}
J{\'o}zsef Fiser and Richard~N Aslin.
\newblock Statistical learning of higher-order temporal structure from visual
  shape sequences.
\newblock \emph{J. Exp. Psychol.}, 28\penalty0 (3):\penalty0 458, 2002.

\bibitem[Steyvers and Tenenbaum(2005)]{Steyvers-01}
Mark Steyvers and Joshua~B Tenenbaum.
\newblock The large-scale structure of semantic networks: {S}tatistical
  analyses and a model of semantic growth.
\newblock \emph{Cogn. Sci.}, 29\penalty0 (1):\penalty0 41--78, 2005.

\bibitem[Dorogovtsev and Mendes(2001)]{Dorogovtsev-01}
Sergey~N Dorogovtsev and Jos{\'e} Fernando~F Mendes.
\newblock Language as an evolving word web.
\newblock \emph{Philos. Trans. R. Soc. Lond., B, Biol. Sci.}, 268\penalty0
  (1485):\penalty0 2603--2606, 2001.

\bibitem[Cancho and Sol{\'e}(2001)]{Cancho-01}
Ramon Ferrer~I Cancho and Richard~V Sol{\'e}.
\newblock The small world of human language.
\newblock \emph{Proc. R. Soc. Lond., B, Biol. Sci.}, 268\penalty0
  (1482):\penalty0 2261--2265, 2001.

\bibitem[Liu et~al.(2010)Liu, Chi, and Small]{Liu-02}
Xiao~Fan Liu, K~Tse Chi, and Michael Small.
\newblock Complex network structure of musical compositions: {A}lgorithmic
  generation of appealing music.
\newblock \emph{Physica A}, 389\penalty0 (1):\penalty0 126--132, 2010.

\bibitem[Barab{\'a}si et~al.(2002)Barab{\'a}si, Jeong, N{\'e}da, Ravasz,
  Schubert, and Vicsek]{Barabasi-02}
Albert-Laszlo Barab{\'a}si, Hawoong Jeong, Zoltan N{\'e}da, Erzsebet Ravasz,
  Andras Schubert, and Tamas Vicsek.
\newblock Evolution of the social network of scientific collaborations.
\newblock \emph{Physica A}, 311\penalty0 (3-4):\penalty0 590--614, 2002.

\bibitem[Girvan and Newman(2002)]{Girvan-01}
Michelle Girvan and Mark~EJ Newman.
\newblock Community structure in social and biological networks.
\newblock \emph{Proc. Natl. Acad. Sci. U.S.A.}, 99\penalty0 (12):\penalty0
  7821--7826, 2002.

\bibitem[Eriksen et~al.(2003)Eriksen, Simonsen, Maslov, and
  Sneppen]{Eriksen-01}
Kasper~Astrup Eriksen, Ingve Simonsen, Sergei Maslov, and Kim Sneppen.
\newblock Modularity and extreme edges of the internet.
\newblock \emph{Phys. Rev. Lett.}, 90\penalty0 (14):\penalty0 148701, 2003.

\bibitem[Albert et~al.(1999)Albert, Jeong, and Barab{\'a}si]{Albert-03}
R{\'e}ka Albert, Hawoong Jeong, and Albert-L{\'a}szl{\'o} Barab{\'a}si.
\newblock Internet: {D}iameter of the world-wide web.
\newblock \emph{Nature}, 401\penalty0 (6749):\penalty0 130, 1999.

\bibitem[Newman(2001)]{Newman-02}
Mark~EJ Newman.
\newblock The structure of scientific collaboration networks.
\newblock \emph{Proc. Natl. Acad. Sci.}, 98\penalty0 (2):\penalty0 404--409,
  2001.

\bibitem[Martin et~al.(2013)Martin, Ball, Karrer, and Newman]{Martin-01}
Travis Martin, Brian Ball, Brian Karrer, and MEJ Newman.
\newblock Coauthorship and citation patterns in the {P}hysical {R}eview.
\newblock \emph{Phys. Rev. E}, 88\penalty0 (1):\penalty0 012814, 2013.

\bibitem[Beckage et~al.(2011)Beckage, Smith, and Hills]{Beckage-01}
Nicole Beckage, Linda Smith, and Thomas Hills.
\newblock Small worlds and semantic network growth in typical and late talkers.
\newblock \emph{PloS One}, 6\penalty0 (5):\penalty0 e19348, 2011.

\bibitem[Dubossarsky et~al.(2017)Dubossarsky, De~Deyne, and
  Hills]{Dubossarsky-01}
Haim Dubossarsky, Simon De~Deyne, and Thomas~T Hills.
\newblock Quantifying the structure of free association networks across the
  life span.
\newblock \emph{Dev. Psychol.}, 53\penalty0 (8):\penalty0 1560, 2017.

\bibitem[Koechlin and Hyafil(2007)]{Koechlin-01}
Etienne Koechlin and Alexandre Hyafil.
\newblock Anterior prefrontal function and the limits of human decision-making.
\newblock \emph{Science}, 318\penalty0 (5850):\penalty0 594--598, 2007.

\bibitem[Wilensky(1983)]{Wilensky-01}
Robert Wilensky.
\newblock Planning and understanding: {A} computational approach to human
  reasoning.
\newblock 1983.

\bibitem[Brent and Cartwright(1996)]{Brent-01}
Michael~R Brent and Timothy~A Cartwright.
\newblock Distributional regularity and phonotactic constraints are useful for
  segmentation.
\newblock \emph{Cognition}, 61\penalty0 (1-2):\penalty0 93--125, 1996.

\bibitem[Romberg and Saffran(2010)]{Romberg-01}
Alexa~R Romberg and Jenny~R Saffran.
\newblock Statistical learning and language acquisition.
\newblock \emph{Wiley Interdiscip. Rev. Cogn. Sci.}, 1\penalty0 (6):\penalty0
  906--914, 2010.

\bibitem[Aslin and Newport(2012)]{Aslin-01}
Richard~N Aslin and Elissa~L Newport.
\newblock Statistical learning: {F}rom acquiring specific items to forming
  general rules.
\newblock \emph{Curr. Dir. Psychol. Sci.}, 21\penalty0 (3):\penalty0 170--176,
  2012.

\bibitem[Aslin and Newport(2014)]{Aslin-02}
Richard~N Aslin and Elissa~L Newport.
\newblock Distributional language learning: {M}echanisms and models of category
  formation.
\newblock \emph{Lang. Learn.}, 64\penalty0 (s2):\penalty0 86--105, 2014.

\bibitem[Schapiro and Turk-Browne(2015)]{Schapiro-02}
A~Schapiro and Nicholas Turk-Browne.
\newblock Statistical learning.
\newblock In \emph{Brain Mapping: An Encyclopedic Reference}, pages 501--506.
  Elsevier, 2015.

\bibitem[Turk-Browne et~al.(2008)Turk-Browne, Isola, Scholl, and
  Treat]{Turk-02}
Nicholas~B Turk-Browne, Phillip~J Isola, Brian~J Scholl, and Teresa~A Treat.
\newblock Multidimensional visual statistical learning.
\newblock \emph{J. Exp. Psychol. Learn. Mem. Cogn.}, 34\penalty0 (2):\penalty0
  399, 2008.

\bibitem[Brady and Oliva(2008)]{Brady-01}
Timothy~F Brady and Aude Oliva.
\newblock Statistical learning using real-world scenes: {E}xtracting
  categorical regularities without conscious intent.
\newblock \emph{Psychol. Sci.}, 19\penalty0 (7):\penalty0 678--685, 2008.

\bibitem[Baldwin et~al.(2008)Baldwin, Andersson, Saffran, and
  Meyer]{Baldwin-01}
Dare Baldwin, Annika Andersson, Jenny Saffran, and Meredith Meyer.
\newblock Segmenting dynamic human action via statistical structure.
\newblock \emph{Cognition}, 106\penalty0 (3):\penalty0 1382--1407, 2008.

\bibitem[Karuza et~al.(2016)Karuza, Thompson-Schill, and Bassett]{Karuza-01}
Elisabeth~A Karuza, Sharon~L Thompson-Schill, and Danielle~S Bassett.
\newblock Local patterns to global architectures: {I}nfluences of network
  topology on human learning.
\newblock \emph{Trends Cogn. Sci.}, 20\penalty0 (8):\penalty0 629--640, 2016.

\bibitem[Newman(2003)]{Newman-01}
Mark~EJ Newman.
\newblock The structure and function of complex networks.
\newblock \emph{SIAM Rev.}, 45\penalty0 (2):\penalty0 167--256, 2003.

\bibitem[Schapiro et~al.(2013)Schapiro, Rogers, Cordova, Turk-Browne, and
  Botvinick]{Schapiro-01}
Anna~C Schapiro, Timothy~T Rogers, Natalia~I Cordova, Nicholas~B Turk-Browne,
  and Matthew~M Botvinick.
\newblock Neural representations of events arise from temporal community
  structure.
\newblock \emph{Nat. Neurosci.}, 16\penalty0 (4):\penalty0 486--492, 2013.

\bibitem[Kahn et~al.(2018)Kahn, Karuza, Vettel, and Bassett]{Kahn-01}
Ari~E Kahn, Elisabeth~A Karuza, Jean~M Vettel, and Danielle~S Bassett.
\newblock Network constraints on learnability of probabilistic motor sequences.
\newblock \emph{Nat. Hum. Behav.}, 2\penalty0 (12):\penalty0 936, 2018.

\bibitem[Lynn et~al.(2019)Lynn, Kahn, and Bassett]{Lynn-06}
Christopher~W Lynn, Ari~E Kahn, and Danielle~S Bassett.
\newblock Structure from noise: {M}ental errors yield abstract representations
  of events.
\newblock \emph{In press}, 2019.

\bibitem[Garvert et~al.(2017)Garvert, Dolan, and Behrens]{Garvert-01}
Mona~M Garvert, Raymond~J Dolan, and Timothy~EJ Behrens.
\newblock A map of abstract relational knowledge in the human
  hippocampal--entorhinal cortex.
\newblock \emph{Elife}, 6, 2017.

\bibitem[Reber(1967)]{Reber-01}
Arthur~S Reber.
\newblock Implicit learning of artificial grammars.
\newblock \emph{J. Verbal Learning Verbal Behav.}, 6\penalty0 (6):\penalty0
  855--863, 1967.

\bibitem[Cleeremans and McClelland(1991)]{Cleeremans-01}
Axel Cleeremans and James~L McClelland.
\newblock Learning the structure of event sequences.
\newblock \emph{J. Exp. Psychol. Gen.}, 120\penalty0 (3):\penalty0 235, 1991.

\bibitem[Gomez and Gerken(1999)]{Gomez-02}
Rebecca~L Gomez and LouAnn Gerken.
\newblock Artificial grammar learning by 1-year-olds leads to specific and
  abstract knowledge.
\newblock \emph{Cognition}, 70\penalty0 (2):\penalty0 109--135, 1999.

\bibitem[Adelman et~al.(2006)Adelman, Brown, and Quesada]{Adelman-01}
James~S Adelman, Gordon~DA Brown, and Jos{\'e}~F Quesada.
\newblock Contextual diversity, not word frequency, determines word-naming and
  lexical decision times.
\newblock \emph{Psychol. Sci.}, 17\penalty0 (9):\penalty0 814--823, 2006.

\bibitem[Balota et~al.(2004)Balota, Cortese, Sergent-Marshall, Spieler, and
  Yap]{Balota-01}
David~A Balota, Michael~J Cortese, Susan~D Sergent-Marshall, Daniel~H Spieler,
  and Melvin~J Yap.
\newblock Visual word recognition of single-syllable words.
\newblock \emph{J. Exp. Psychol.}, 133\penalty0 (2):\penalty0 283, 2004.

\bibitem[Carlson et~al.(2014)Carlson, Sonderegger, and Bane]{Carlson-01}
Matthew~T Carlson, Morgan Sonderegger, and Max Bane.
\newblock How children explore the phonological network in child-directed
  speech: {A} survival analysis of children's first word productions.
\newblock \emph{J. Mem. Lang.}, 75:\penalty0 159--180, 2014.

\bibitem[Chan and Vitevitch(2009)]{Chan-01}
Kit~Ying Chan and Michael~S Vitevitch.
\newblock The influence of the phonological neighborhood clustering coefficient
  on spoken word recognition.
\newblock \emph{J. Exp. Psychol. Hum. Percept. Perform.}, 35\penalty0
  (6):\penalty0 1934, 2009.

\bibitem[Goldstein and Vitevitch(2014)]{Goldstein-01}
Rutherford Goldstein and Michael~S Vitevitch.
\newblock The influence of clustering coefficient on word-learning: {H}ow
  groups of similar sounding words facilitate acquisition.
\newblock \emph{Front. Psychol.}, 5:\penalty0 1307, 2014.

\bibitem[Yates(2013)]{Yates-01}
Mark Yates.
\newblock How the clustering of phonological neighbors affects visual word
  recognition.
\newblock \emph{J. Exp. Psychol. Learn. Mem. Cogn.}, 39\penalty0 (5):\penalty0
  1649, 2013.

\bibitem[Masucci et~al.(2011)Masucci, Kalampokis, Egu{\'\i}luz, and
  Hern{\'a}ndez-Garc{\'\i}a]{Masucci-01}
Adolfo~Paolo Masucci, Alkiviadis Kalampokis, Victor~Mart{\'\i}nez Egu{\'\i}luz,
  and Emilio Hern{\'a}ndez-Garc{\'\i}a.
\newblock Wikipedia information flow analysis reveals the scale-free
  architecture of the semantic space.
\newblock \emph{PloS One}, 6\penalty0 (2):\penalty0 e17333, 2011.

\bibitem[Barab{\'a}si and Albert(1999)]{Barabasi-01}
Albert-L{\'a}szl{\'o} Barab{\'a}si and R{\'e}ka Albert.
\newblock Emergence of scaling in random networks.
\newblock \emph{Science}, 286\penalty0 (5439):\penalty0 509--512, 1999.

\bibitem[Borge-Holthoefer and Arenas(2010)]{Borge-01}
Javier Borge-Holthoefer and Alex Arenas.
\newblock Semantic networks: {S}tructure and dynamics.
\newblock \emph{Entropy}, 12\penalty0 (5):\penalty0 1264--1302, 2010.

\bibitem[Karuza et~al.(2017)Karuza, Kahn, Thompson-Schill, and
  Bassett]{Karuza-02}
Elisabeth~A Karuza, Ari~E Kahn, Sharon~L Thompson-Schill, and Danielle~S
  Bassett.
\newblock Process reveals structure: {H}ow a network is traversed mediates
  expectations about its architecture.
\newblock \emph{Sci. Rep.}, 7\penalty0 (1):\penalty0 12733, 2017.

\bibitem[Karuza et~al.(2019)Karuza, Kahn, and Bassett]{Karuza-03}
Elisabeth~A Karuza, Ari~E Kahn, and Danielle~S Bassett.
\newblock Human sensitivity to community structure is robust to topological
  variation.
\newblock \emph{Complexity}, 2019, 2019.

\bibitem[Hyman(1953)]{Hyman-01}
Ray Hyman.
\newblock Stimulus information as a determinant of reaction time.
\newblock \emph{J. Exp. Psychol.}, 45\penalty0 (3):\penalty0 188, 1953.

\bibitem[McCarthy and Donchin(1981)]{McCarthy-01}
Gregory McCarthy and Emanuel Donchin.
\newblock A metric for thought: {A} comparison of p300 latency and reaction
  time.
\newblock \emph{Science}, 211\penalty0 (4477):\penalty0 77--80, 1981.

\bibitem[Forster and Chambers(1973)]{Forster-01}
Kenneth~I Forster and Susan~M Chambers.
\newblock Lexical access and naming time.
\newblock \emph{J. Verbal Learning Verbal Behav.}, 12\penalty0 (6):\penalty0
  627--635, 1973.

\bibitem[Anderson(1974)]{Anderson-01}
John~Robert Anderson.
\newblock Retrieval of propositional information from long-term memory.
\newblock \emph{Cogn. Psychol.}, 6\penalty0 (4):\penalty0 451--474, 1974.

\bibitem[Anderson and Reder(1999)]{Anderson-02}
John~R Anderson and Lynne~M Reder.
\newblock The fan effect: {N}ew results and new theories.
\newblock \emph{J. Exp. Psychol.}, 128\penalty0 (2):\penalty0 186, 1999.

\bibitem[Motter et~al.(2002)Motter, De~Moura, Lai, and Dasgupta]{Motter-01}
Adilson~E Motter, Alessandro~PS De~Moura, Ying-Cheng Lai, and Partha Dasgupta.
\newblock Topology of the conceptual network of language.
\newblock \emph{Phys. Rev. E}, 65\penalty0 (6):\penalty0 065102, 2002.

\bibitem[Sigman and Cecchi(2002)]{Sigman-01}
Mariano Sigman and Guillermo~A Cecchi.
\newblock Global organization of the wordnet lexicon.
\newblock \emph{Proc. Natl. Acd. Sci.}, 99\penalty0 (3):\penalty0 1742--1747,
  2002.

\bibitem[Watts and Strogatz(1998)]{Watts-01}
Duncan~J Watts and Steven~H Strogatz.
\newblock Collective dynamics of `small-world' networks.
\newblock \emph{Nature}, 393\penalty0 (6684):\penalty0 440, 1998.

\bibitem[Newman(2006)]{Newman-03}
Mark~EJ Newman.
\newblock Modularity and community structure in networks.
\newblock \emph{Proc. Natl. Acad. Sci.}, 103\penalty0 (23):\penalty0
  8577--8582, 2006.

\bibitem[Moody(2001)]{Moody-01}
James Moody.
\newblock Peer influence groups: {I}dentifying dense clusters in large
  networks.
\newblock \emph{Soc. Netw.}, 23\penalty0 (4):\penalty0 261--283, 2001.

\bibitem[Vitevitch et~al.(2012)Vitevitch, Chan, and Roodenrys]{Vitevitch-01}
Michael~S Vitevitch, Kit~Ying Chan, and Steven Roodenrys.
\newblock Complex network structure influences processing in long-term and
  short-term memory.
\newblock \emph{J. Mem. Lang.}, 67\penalty0 (1):\penalty0 30--44, 2012.

\bibitem[Chan and Vitevitch(2010)]{Chan-02}
Kit~Ying Chan and Michael~S Vitevitch.
\newblock Network structure influences speech production.
\newblock \emph{Cogn. Sci.}, 34\penalty0 (4):\penalty0 685--697, 2010.

\bibitem[Kleinberg(2000)]{Kleinberg-01}
Jon~M Kleinberg.
\newblock Navigation in a small world.
\newblock \emph{Nature}, 406\penalty0 (6798):\penalty0 845, 2000.

\bibitem[Boas(2006)]{Boas-01}
Mary~L Boas.
\newblock \emph{Mathematical methods in the physical sciences}.
\newblock Wiley, 2006.

\bibitem[Howard and Kahana(2002)]{Howard-01}
Marc~W Howard and Michael~J Kahana.
\newblock A distributed representation of temporal context.
\newblock \emph{J. Math. Psychol.}, 46\penalty0 (3):\penalty0 269--299, 2002.

\bibitem[Dehaene et~al.(2015)Dehaene, Meyniel, Wacongne, Wang, and
  Pallier]{Dehaene-01}
Stanislas Dehaene, Florent Meyniel, Catherine Wacongne, Liping Wang, and
  Christophe Pallier.
\newblock The neural representation of sequences: {F}rom transition
  probabilities to algebraic patterns and linguistic trees.
\newblock \emph{Neuron}, 88\penalty0 (1):\penalty0 2--19, 2015.

\bibitem[Meyniel and Dehaene(2017)]{Meyniel-01}
Florent Meyniel and Stanislas Dehaene.
\newblock Brain networks for confidence weighting and hierarchical inference
  during probabilistic learning.
\newblock \emph{Proc. Natl. Acad. Sci. U.S.A.}, page 201615773, 2017.

\bibitem[Angela and Cohen(2009)]{Angela-01}
J~Yu Angela and Jonathan~D Cohen.
\newblock Sequential effects: {S}uperstition or rational behavior?
\newblock In \emph{Advances in Neural Information Processing Systems}, pages
  1873--1880, 2009.

\bibitem[Meyniel et~al.(2016)Meyniel, Maheu, and Dehaene]{Meyniel-02}
Florent Meyniel, Maxime Maheu, and Stanislas Dehaene.
\newblock Human inferences about sequences: {A} minimal transition probability
  model.
\newblock \emph{PLOS Comput. Biol.}, 12\penalty0 (12):\penalty0 e1005260, 2016.

\bibitem[Momennejad et~al.(2017)Momennejad, Russek, Cheong, Botvinick, Daw, and
  Gershman]{Momennejad-01}
Ida Momennejad, Evan~M Russek, Jin~H Cheong, Matthew~M Botvinick, ND~Daw, and
  Samuel~J Gershman.
\newblock The successor representation in human reinforcement learning.
\newblock \emph{Nat. Hum. Behav.}, 1\penalty0 (9):\penalty0 680, 2017.

\bibitem[Newport and Aslin(2004)]{Newport-01}
Elissa~L Newport and Richard~N Aslin.
\newblock Learning at a distance {I}. {S}tatistical learning of non-adjacent
  dependencies.
\newblock \emph{Cogn. Psychol.}, 48\penalty0 (2):\penalty0 127--162, 2004.

\bibitem[Sutton et~al.(1998)Sutton, Barto, et~al.]{Sutton-02}
Richard~S Sutton, Andrew~G Barto, et~al.
\newblock \emph{Introduction to reinforcement learning}, volume~2.
\newblock MIT press Cambridge, 1998.

\bibitem[Dayan(1993)]{Dayan-01}
Peter Dayan.
\newblock Improving generalization for temporal difference learning: {T}he
  successor representation.
\newblock \emph{Neural Comput.}, 5\penalty0 (4):\penalty0 613--624, 1993.

\bibitem[Gershman et~al.(2012)Gershman, Moore, Todd, Norman, and
  Sederberg]{Gershman-01}
Samuel~J Gershman, Christopher~D Moore, Michael~T Todd, Kenneth~A Norman, and
  Per~B Sederberg.
\newblock The successor representation and temporal context.
\newblock \emph{Neural Comput.}, 24\penalty0 (6):\penalty0 1553--1568, 2012.

\bibitem[Altmann and Kamide(1999)]{Altmann-01}
Gerry~TM Altmann and Yuki Kamide.
\newblock Incremental interpretation at verbs: {R}estricting the domain of
  subsequent reference.
\newblock \emph{Cognition}, 73\penalty0 (3):\penalty0 247--264, 1999.

\bibitem[Atance and O'Neill(2001)]{Atance-01}
Cristina~M Atance and Daniela~K O'Neill.
\newblock Episodic future thinking.
\newblock \emph{Trends Cogn. Sci.}, 5\penalty0 (12):\penalty0 533--539, 2001.

\bibitem[Addis et~al.(2008)Addis, Wong, and Schacter]{Addis-01}
Donna~Rose Addis, Alana~T Wong, and Daniel~L Schacter.
\newblock Age-related changes in the episodic simulation of future events.
\newblock \emph{Psychol. Sci.}, 19\penalty0 (1):\penalty0 33--41, 2008.

\bibitem[Huettel et~al.(2002)Huettel, Mack, and McCarthy]{Huettel-01}
Scott~A Huettel, Peter~B Mack, and Gregory McCarthy.
\newblock Perceiving patterns in random series: dynamic processing of sequence
  in prefrontal cortex.
\newblock \emph{Nat. Neurosci.}, 5\penalty0 (5):\penalty0 485, 2002.

\bibitem[Falk and Konold(1997)]{Falk-01}
Ruma Falk and Clifford Konold.
\newblock Making sense of randomness: Implicit encoding as a basis for
  judgment.
\newblock \emph{Psychol. Rev.}, 104\penalty0 (2):\penalty0 301, 1997.

\bibitem[Squires et~al.(1976)Squires, Wickens, Squires, and
  Donchin]{Squires-01}
Kenneth~C Squires, Christopher Wickens, Nancy~K Squires, and Emanuel Donchin.
\newblock The effect of stimulus sequence on the waveform of the cortical
  event-related potential.
\newblock \emph{Science}, 193\penalty0 (4258):\penalty0 1142--1146, 1976.

\bibitem[Ross et~al.(1996)Ross, Kelly, Sullivan, Perry, Mercer, Davis,
  Washburn, Sager, Boyce, and Bristow]{Ross-01}
Sheldon~M Ross, John~J Kelly, Roger~J Sullivan, William~James Perry, Donald
  Mercer, Ruth~M Davis, Thomas~Dell Washburn, Earl~V Sager, Joseph~B Boyce, and
  Vincent~L Bristow.
\newblock \emph{Stochastic Processes}, volume~2.
\newblock Wiley New York, 1996.

\bibitem[Amit et~al.(1994)Amit, Shmerler, Eisenberg, Abraham, and
  Shnerb]{Amit-01}
M~Amit, Y~Shmerler, E~Eisenberg, M~Abraham, and N~Shnerb.
\newblock Language and codification dependence of long-range correlations in
  texts.
\newblock \emph{Fractals}, 2\penalty0 (01):\penalty0 7--13, 1994.

\bibitem[Jafari et~al.(2007)Jafari, Pedram, and Hedayatifar]{Jafari-01}
GR~Jafari, P~Pedram, and L~Hedayatifar.
\newblock Long-range correlation and multifractality in {B}ach's inventions
  pitches.
\newblock \emph{J. Stat. Mech.: Theory Exp.}, \penalty0 (04), 2007.

\bibitem[Adamic et~al.(2001)Adamic, Lukose, Puniyani, and Huberman]{Adamic-02}
Lada~A Adamic, Rajan~M Lukose, Amit~R Puniyani, and Bernardo~A Huberman.
\newblock Search in power-law networks.
\newblock \emph{Phys. Rev. E}, 64\penalty0 (4):\penalty0 046135, 2001.

\bibitem[Dodds et~al.(2003)Dodds, Muhamad, and Watts]{Dodds-01}
Peter~Sheridan Dodds, Roby Muhamad, and Duncan~J Watts.
\newblock An experimental study of search in global social networks.
\newblock \emph{Science}, 301\penalty0 (5634):\penalty0 827--829, 2003.

\bibitem[O'Day and Jeffries(1993)]{Oday-01}
Vicki~L O'Day and Robin Jeffries.
\newblock Orienteering in an information landscape: {H}ow information seekers
  get from here to there.
\newblock In \emph{Proceedings of the INTERACT'93 and CHI'93 conference on
  Human factors in computing systems}, pages 438--445. ACM, 1993.

\bibitem[West and Leskovec(2012)]{West-01}
Robert West and Jure Leskovec.
\newblock Human wayfinding in information networks.
\newblock In \emph{Proceedings of the 21st international conference on World
  Wide Web}, pages 619--628. ACM, 2012.

\bibitem[Raaijmakers and Shiffrin(1981)]{Raaijmakers-02}
Jeroen~G Raaijmakers and Richard~M Shiffrin.
\newblock Search of associative memory.
\newblock \emph{Psychol. Rev.}, 88\penalty0 (2):\penalty0 93, 1981.

\bibitem[Hills et~al.(2012)Hills, Jones, and Todd]{Hills-02}
Thomas~T Hills, Michael~N Jones, and Peter~M Todd.
\newblock Optimal foraging in semantic memory.
\newblock \emph{Psychol. Rev.}, 119\penalty0 (2):\penalty0 431, 2012.

\bibitem[Jones et~al.(2015)Jones, Hills, and Todd]{Jones-01}
MN~Jones, TT~Hills, and PM~Todd.
\newblock Hidden processes in structural representations: {A} reply to abbott,
  austerweil, and griffiths (2015).
\newblock \emph{Psychol. Rev.}, 122\penalty0 (3):\penalty0 570--574, 2015.

\bibitem[Avery and Jones(2018)]{Avery-01}
Johnathan Avery and Michael~N Jones.
\newblock Comparing models of semantic fluency: {D}o humans forage optimally,
  or walk randomly?
\newblock In \emph{Proceedings of the 40th Annual Meeting of the 40th Annual
  Meeting of the Cognitive Science Society}, 2018.

\bibitem[Schulz-Hardt et~al.(2000)Schulz-Hardt, Frey, L{\"u}thgens, and
  Moscovici]{Schulz-01}
Stefan Schulz-Hardt, Dieter Frey, Carsten L{\"u}thgens, and Serge Moscovici.
\newblock Biased information search in group decision making.
\newblock \emph{J. Pers. Soc. Psychol.}, 78\penalty0 (4):\penalty0 655, 2000.

\bibitem[Jonas et~al.(2001)Jonas, Schulz-Hardt, Frey, and Thelen]{Jonas-01}
Eva Jonas, Stefan Schulz-Hardt, Dieter Frey, and Norman Thelen.
\newblock Confirmation bias in sequential information search after preliminary
  decisions: {A}n expansion of dissonance theoretical research on selective
  exposure to information.
\newblock \emph{J. Pers. Soc. Psychol.}, 80\penalty0 (4):\penalty0 557, 2001.

\bibitem[Ravasz and Barab{\'a}si(2003)]{Ravasz-01}
Erzs{\'e}bet Ravasz and Albert-L{\'a}szl{\'o} Barab{\'a}si.
\newblock Hierarchical organization in complex networks.
\newblock \emph{Phys. Rev. E}, 67\penalty0 (2):\penalty0 026112, 2003.

\bibitem[Arenas et~al.(2008)Arenas, Fernandez, and Gomez]{Arenas-02}
Alex Arenas, Alberto Fernandez, and Sergio Gomez.
\newblock Analysis of the structure of complex networks at different resolution
  levels.
\newblock \emph{New J. Phys.}, 10\penalty0 (5):\penalty0 053039, 2008.

\end{thebibliography}

% Supplement:
\newpage

\begin{figure*}[t]
\centering
\includegraphics[width=.85\linewidth]{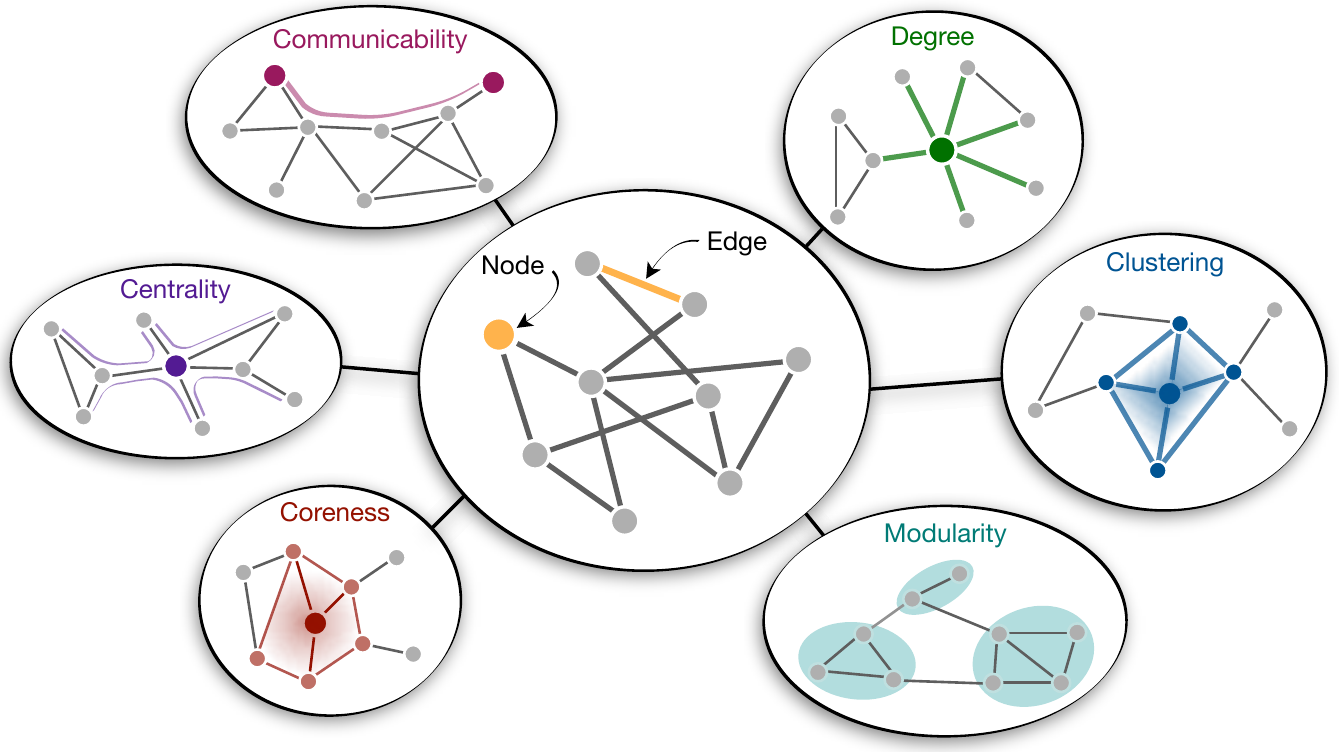}
\captionsetup{labelformat=empty}
\caption{\textbf{Supplementary Fig. 1.} A primer on network properties. (\emph{Center}) Nodes, illustrated by circles, represent stimuli, items, or states in a sequence. Edges, illustrated by lines, connect pairs of nodes if it is possible to transition from one node to the other. The organization of edges among nodes is referred to as the network's \emph{topology} or \emph{structure}. (\emph{Circumjacent}) A network's topology can be described using properties that characterize its local, mesoscale, or global organization. For example, the simplest local property is the degree of a node (green), or the number of edges emanating from a node. Two notions of mesoscale structure include (i) the clustering coefficient (blue), or the ratio of connected triangles to connected triples of nodes, and (ii) modularity (turquoise), where there exist communities of nodes with internally dense and externally sparse connections. Finally, global measures include (i) coreness (red), or the ability of a node to withstand the removal of nodes with low degree, (ii) notions of centrality (purple) such as betweenness centrality, which quantifies the importance of a node for facilitating long-distance connections, and (iii) communicability (magenta), which captures the number of paths of various lengths connecting two nodes. Collectively, the network representation and associated properties can provide critical insights into the structure of the system under study.}
\label{fig_primer}
\end{figure*}

\end{document}